\documentclass[12pt]{iopart}

\usepackage{graphicx}
\usepackage{dcolumn}
\usepackage{bm}
\usepackage{psfrag}
\usepackage{subfigure}

\newcommand{\be}{\begin{eqnarray}}
\newcommand{\ee}{\end{eqnarray}}
\newcommand{\bn}{\begin{eqnarray*}}
\newcommand{\en}{\end{eqnarray*}}

\newcommand{\nn}{\nonumber \\}
\newcommand{\nl}{\\}

\renewcommand{\vec}[1]{\mbox{\boldmath$#1$}}
\renewcommand{\d}{\mbox{\rm d}}

\newcommand{\ph}{\ensuremath{\phi}}
\newcommand{\vph}{\ensuremath{\varphi}}
\newcommand{\al}{\ensuremath{\alpha}}
\newcommand{\bt}{\ensuremath{\beta}}

\newcommand{\gm}{\ensuremath{\gamma}}
\newcommand{\dl}{\ensuremath{\delta}}
\newcommand{\lm}{\ensuremath{\lambda}}

\newcommand{\Dl}{\ensuremath{\Delta}}

\newcommand{\pvec}{\ensuremath{\vec{p}}}
\newcommand{\Pvec}{\ensuremath{\vec{P}}}

\newcommand{\xivec}{\ensuremath{\vec{\xi}}}
\newcommand{\xvec}{\ensuremath{\vec{x}}}

\newcommand{\Jvec}{\ensuremath{\vec{J}}}
\newcommand{\Lvec}{\ensuremath{\vec{L}}}
\newcommand{\Svec}{\ensuremath{\vec{S}}}

\newcommand{\Xvec}{\ensuremath{\vec{X}}}
\newcommand{\Yvec}{\ensuremath{\vec{Y}}}
\newcommand{\Zvec}{\ensuremath{\vec{Z}}}

\newcommand{\Omvec}{\ensuremath{\vec{\Omega}}}

\newcommand{\nabvec}{\ensuremath{\vec{\nabla}}}

\newcommand{\hb}{\ensuremath{\hbar}}

\newcommand{\lt}{\ensuremath{\left}}
\newcommand{\rt}{\ensuremath{\right}}

\renewcommand{\d}{\mbox{\rm d}}

\begin{document}

\title[Local Space-Time Curvature Effects on Quantum Orbital Angular Momentum]
{Local Space-Time Curvature Effects on Quantum Orbital Angular Momentum}

\author{Dinesh Singh\dag
\footnote[3]{Author to whom correspondence should be addressed.} \, and
Nader Mobed\dag
}

\address{\dag\ Department of Physics, University of Regina \\
Regina, Saskatchewan, S4S 0A2, Canada}


\eads{\mailto{dinesh.singh@uregina.ca} and \mailto{nader.mobed@uregina.ca}
}

\begin{abstract}
This paper claims that local space-time curvature can non-trivially contribute to the properties of
orbital angular momentum in quantum mechanics.
Of key importance is the demonstration that an extended orbital angular momentum operator due to gravitation
can identify the existence of orbital states with half-integer projection quantum numbers $m$
along the axis of quantization, while still preserving integer-valued orbital quantum numbers $l$
for a simply connected topology.
The consequences of this possibility are explored in depth, noting that the half-integer $m$ states
vanish as required when the locally curved space-time reduces to flat space-time,
fully recovering all established properties of orbital angular momentum in this limit.
In particular, it is shown that a minimum orbital number of $l = 2$ is necessary for the gravitational
interaction to appear within this context, in perfect correspondence with the spin-2 nature of
linearized general relativity.
\end{abstract}




\submitto{\CQG}



\section{Introduction}

In the search for a self-consistent quantum theory of gravity over the past seventy years or more,
there now exist many distinct avenues available to pursue.
Arguably, the leading contenders in this pursuit are string theory \cite{Polchinski,Green} and
loop quantum gravity \cite{Ashtekar,Rovelli}.
This is followed---to varying degrees of interest---by twistor theory \cite{Penrose},
causal set theory \cite{Sorkin}, Regge calculus and causal dynamical triangulations \cite{Williams,Loll},
noncommutative geometry \cite{Doplicher}, and so forth.
In the absence of any discernable physical evidence to offer guidance towards achieving this ultimate goal,
these and other competing theories are mathematically very sophisticated,
requiring a great deal of ingenuity and intensive effort in order to make any headway along any one
of these directions.
In spite of their differences on a wide array of conceptual and computational details,
all these approaches effectively claim that some form of
unification involving gravitation and quantum mechanics becomes relevant only at the Planck length scale of
\mbox{$10^{-33}$ cm}, some twenty orders of magnitude smaller than the effective radius of a proton.
Addressing this basic discrepancy of scale makes it extremely difficult to conceive of realistic tests for their
efficacy, with the possible exception of identifying cosmologically-driven observations to amplify any
potential quantum gravity signatures to length scales large enough for detection \cite{Amelino-Camelia}.

Given this dilemma, it is worthwhile to ask if there are more indirect and modest means to seek out
quantum gravity, but strongly driven by a simple desire to acquire some readily identifiable predictions
involving established theories of gravitation and quantum mechanics when put under extreme conditions.
In other words, is it possible to identify a suitable length scale many orders of magnitude higher
than the Planck scale, in which general relativity and quantum mechanics may overlap and interact
in unforeseen ways, such that a realistic possibility for experimental tests can be theorized?
To this question, an affirmative answer exists, motivated largely by an exploration of known
phenomena whereby certain critical assumptions are isolated and scrutinized in depth.

For example, in large part to address the so-called {\it hypothesis of locality} \cite{Mashhoon}
when applied to quantum mechanical particles while in non-inertial motion
or gravitationally accelerated \cite{Singh-Mobed-Papini-helicity},
it is shown that Casimir invariance for spin, as described by the Poincar\'{e} group, no longer holds true
\cite{Singh-Mobed-Casimir-1,Singh-Mobed-Casimir-2,Singh-Mobed-Casimir-3}, and that
this breakdown of formalism results in readily identifiable and interesting physical predictions
that may be potentially observable.
Such a predicted breakdown can also be compared with other approaches \cite{Kosinski} to the deformation
of the Poincar\'{e} group, whose Casimir operators for mass and spin can be determined.
A second example concerns the interaction of neutrinos in a curved space-time background, in which
it is shown that a single neutrino described as a wave packet is sensitive to variations of
space-time curvature \cite{Singh-Mobed-Papini-neutrino-1}, such that Dirac and Majorana
neutrinos can be rendered distinguishable without recourse to hypothetical particle physics
beyond the Standard Model.
In addition, for this model as applied to a two-flavour oscillation,
it is theoretically possible to infer the absolute neutrino masses due to gravitational interactions alone
\cite{Singh-Mobed-Papini-neutrino-1,Singh-Mobed-Papini-neutrino-2}.
For a third example, it is shown that spin-1/2 particle {\it Zitterbewegung} in the presence
of a local gravitational background \cite{Singh-Mobed-zitt} while propagating through space-time
generates a quantum violation of the weak equivalence principle, due to the explicit coupling of
the background Ricci curvature tensor to the particle's mass-dependent {\it Zitterbewegung} frequency,
which is inversely proportional to the particle's Compton wavelength.

With the same underlying motivations as cited above, this paper is intended to present some predicted
physical consequences for the interaction of a spinless quantum mechanical particle
with nonzero orbital angular momentum within a locally curved space-time setting.
It is overwhelmingly evident that, for example, the presence of a terrestrial gravitational background
has for all practical purposes no discernable impact on the local properties of orbital angular momentum for some
valence electron surrounding an atom at rest with respect to its environment.
At the same time, it should also be evident that, by the weak principle of equivalence and
geodesic deviation in classical general relativity, this same electron must still interact with local
tidal acceleration effects at some space-like separation away from a reference frame situated
at the centre of the atom's nucleus.
Since it is impossible to remove the perceived effects of a gravitational background beyond a
mathematical point, in principle they must manifest some predictive consequences for the
properties of orbital angular momentum, no matter how small they may be at a practical level.
While it is reasonable to question whether it is realistic to envision sensing
local gravitational effects for atomic systems, on a purely theoretical level the outcome of this
exploration may have a potentially significant impact on how to sharpen the focus of inquiry
as it pertains to quantum gravity research.

A major finding of this paper is the observation that the orbiting particle is
sensitive to {\em half-integer} spacings along the axis of quantization
when the expectation is to observe strictly {\em integer-valued} spacings only.
Given this conventional wisdom for quantum mechanics, such an observation to the contrary
is very surprising, whose impact is keenly felt throughout this paper.
In retrospect, however, there may already exist some basis in the literature suggesting
that a classical gravitational background can theoretically reveal the presence of half-integer spin angular
momentum states within the context of either nontrivial topological spaces \cite{Friedman,Samuel}
or specific metric configurations \cite{Williams-1}.
Since this paper implicitly assumes a simply connected topology for the space-time,
it follows from the relevant literature \cite{Finkelstein} that only the
prediction of integer-valued spin for the gravitational field is allowed following a $2 \pi$ rotation
of an isolated space-time patch with respect to its environment.
Indeed, for this paper such a requirement is satisfied.
However, a truly significant observation to follow is that the minimum allowable
orbital quantum number for the spinless particle to be sensitive to the gravitational background
is $l = 2$, which precisely matches the spin angular momentum of a graviton, the wider
implications of which are worth exploring in detail.

This paper begins with a physical motivation found in \S \ref{Motivation} to justify the study of
orbital angular momentum in locally curved space-time.
An outline of the basic formalism of orbital angular momentum within standard quantum mechanics immediately follows
in \S \ref{Formalism}, which lays the foundation for presenting in \S \ref{Orbital} the main details
of this paper as it concerns the role of local gravitation on observables and the consequences
that follow.
An in-depth discussion concerning the specifics of this paper, including its possible connections
with existing literature, is then found in \S \ref{Discussion}, with a brief conclusion given in \S \ref{Conclusion}.
For this paper, the space-time metric background is expressed in terms of +2 signature and the conventions
adopted by Misner, Thorne, and Wheeler \cite{MTW}.

\section{Physical Motivation}
\label{Motivation}

It is worthwhile to consider the following physical motivations for this paper.
Suppose that space-time curvature is represented in terms of either Fermi or Riemann normal co-ordinates
$x^\mu = (\tau, \xvec(\tau))$, where $\tau$ is the proper time defined with respect to some reference worldline.
An orthonormal tetrad $\bar{e}^{\hat{\mu}}{}_\nu = \dl^\mu{}_\nu + \tilde{R}^\mu{}_\nu$ is then assumed,
such that hatted indices describe a local Lorentz frame, and $\tilde{R}^\mu{}_\nu$ is a two-indexed
space-time curvature deviation away from the locally flat space-time background, not to be confused with the
Ricci tensor.
Since the local curved space-time metric is
$g_{\mu \nu} = \eta_{\hat{\al} \hat{\bt}} \, \bar{e}^{\hat{\al}}{}_\mu \, \bar{e}^{\hat{\bt}}{}_\nu$,
it follows naturally that \cite{Poisson-1}
\be
{}^F \tilde{R}^\mu{}_\nu & = & \lt[{1 \over 2} \, {}^F R^\mu{}_{lm0} (\tau) \, \dl^0{}_\nu
+ {1 \over 6} \, {}^F R^\mu{}_{lmk} (\tau) \, \dl^k{}_\nu \rt] \dl x^l \, \dl x^m
\nn
\label{FNC-curvature}
\ee
in Fermi normal co-ordinates, and
\be
{}^R \tilde{R}^\mu{}_\nu & = & {1 \over 6} \, {}^R R^\mu{}_{\al \bt \nu} (\tau) \, \dl x^\al \, \dl x^\bt
\label{RNC-curvature}
\ee
in Riemann normal co-ordinates, where $R^\mu{}_{\al \bt \nu} (\tau)$ describes the Riemann curvature tensor
in the local Lorentz frame, and $\dl x^\mu$ is interpreted as a space-time quantum fluctuation
with $|\dl x^\mu| \ll |x^\mu|$ to satisfy $\tilde{R}^\mu{}_\nu \ll \dl^\mu{}_\nu$.

Now suppose that the corresponding position ket vector for normal co-ordinates is described by
$\lt| x^\mu \rt\rangle = \lt| (\tau, \xvec) \rt\rangle$ that is subject to infinitesimal rotation of the spatial
co-ordinates $x^i$ about the origin by small angle $\d \ph$ via a rotation operator
${\cal D}^{(0)}_i (\d \ph)$ \cite{Sakurai}.
Then, for a rotation about a locally defined $z$-direction as the axis of quantization, it follows that
\be
{\cal D}^{(0)}_z (\d \phi) \, \lt| x^\mu \rt\rangle & = & \lt(1 - {i \over \hb} \, \d \ph \, \Lvec^{(0)}_z \rt)
\lt| (\tau, x, y, z) \rt\rangle
\nn
& = & \lt| (\tau, x - y \, \d \ph, y + x \, \d \ph, z) \rt\rangle \, ,
\label{z-rotation-normal-co-ordinates}
\ee
leading to an orbital angular momentum operator
\be
\Lvec^{(0)}_i & = & \varepsilon_{0ijk} \, \xvec^j \, \pvec^k
\label{L0-i}
\ee
defined for normal co-ordinates, where $\varepsilon_{\mu \nu \al \bt}$ is the Levi-Civita tensor density \cite{Carmeli}
with $\varepsilon_{0123} \equiv + 1$.
By following standard arguments involving the quantization of orbital angular momentum \cite{Sakurai}, it follows that
eigenstates $\lt| l, m \rt\rangle$ of $\Lvec^2_{(0)}$ and $\Lvec^{(0)}_z$ exist,
denoted by integers $(l, m)$, such that
\numparts
\be
\Lvec^2_{(0)} \, \lt| l, m \rt\rangle & = & l(l + 1) \, \hb^2 \, \lt| l, m \rt\rangle \, ,
\label{L0^2|lm>}
\nl
\Lvec^{(0)}_z \, \lt| l, m \rt\rangle & = & m \, \hb \, \lt| l, m \rt\rangle \, , \qquad -l \ \leq \ m \ \leq \ l \, .
\label{L0z|lm>}
\ee
\endnumparts

\begin{figure}
\psfrag{tau}[cc][][1.8][0]{\Large $\tau$}
\psfrag{x}[cc][][1.8][0]{\Large $\xvec$}
\psfrag{T}[cc][][1.8][0]{\Large $T$}
\psfrag{X}[cc][][1.8][0]{\Large $\Xvec$}
\psfrag{xi}[cc][][1.8][0]{\Large $\xivec$}
\psfrag{U-proj}[cc][][1.8][0]{\Large $U_{\rm Proj.}(\tau, \xvec)$}
\psfrag{x-ket}[cc][][1.8][0]{\Large $|x^\mu\rangle$}
\psfrag{X-ket}[cc][][1.8][0]{\Large $|\Xvec^{\hat{\mu}} (\tau, \xvec) \rangle_{\rm G}$}
\hspace{2cm}
\begin{minipage}[t]{0.3 \textwidth}
\centering
{
\rotatebox{0}{\includegraphics[width = 8.5cm, height = 6.5cm, scale = 1]{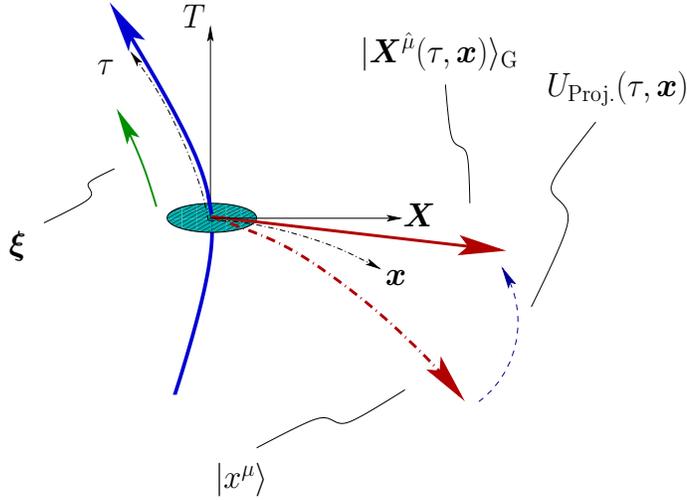}}}
\end{minipage}%
\caption{\label{fig:ket-rotation} $U_{\rm Proj.}(\tau, \xvec)$ is a projection operator to transform the normal
co-ordinate position ket vector $\lt| x^\mu \rt\rangle$ into $|\Xvec^{\hat{\mu}} (\tau, \xvec) \rangle_{\rm G}
= U_{\rm Proj.}(\tau, \xvec) \lt| x^\mu \rt\rangle$ defined with respect to $\tau$.
It is also possible to describe proper time translation of $|\Xvec^{\hat{\mu}} (\tau, \xvec) \rangle_{\rm G}$
in terms of Lie transport with respect to a vector field $\xivec$ tangent to the reference worldline.}
\end{figure}

Consider now the fact that a position vector in a local Lorentz frame is described by
$\Xvec^{\hat{\mu}} (\tau, \xvec) = \bar{e}^{\hat{\mu}}{}_\nu \, x^\nu$, with
$\eta_{\hat{\al} \hat{\bt}} \, \Xvec^{\hat{\al}} \, \Xvec^{\hat{\bt}} = g_{\mu \nu} \, x^\mu \, x^\nu$.
It is also possible to obtain a corresponding position ket vector $|\Xvec^{\hat{\mu}} (\tau, \xvec) \rangle_{\rm G}$
that is unitarily equivalent to $\lt| x^\mu \rt\rangle$, in the form $|\Xvec^{\hat{\mu}} (\tau, \xvec) \rangle_{\rm G} =
\lt| \bar{e}^{\hat{\mu}}{}_\nu \, x^\nu \rt\rangle_{\rm G} = U_{\rm Proj.}(\tau, \xvec) \lt| x^\mu \rt\rangle$,
where 
\be
U_{\rm Proj.}(\tau, \xvec) & = & 1 + \tilde{R}^\bt{}_\al \lt[\xvec^\al \, \nabvec_\bt\rt]_{\rm S}
\ = \
1 - {i \over \hb} \, \tilde{R}_{\bt \al} \lt[\xvec^\al \, \pvec^\bt\rt]_{\rm S}
\label{U-Proj.}
\ee
is a constructed unitary operator to project objects defined in a locally curved space-time in terms of
the local Lorentz frame for each instance of $\tau$.
Symmetrization ``S'' of the position and momentum operators in accordance with Weyl ordering is imposed
to ensure the unitarity of $U_{\rm Proj.}(\tau, \xvec)$.
A visual representation of $U_{\rm Proj.}(\tau, \xvec)$ to illustrate its properties is given by
Figure~\ref{fig:ket-rotation}.

If a unitary transformation of (\ref{L0-i}) is now applied using (\ref{U-Proj.}), it becomes evident that
\be
\Lvec^{(0), {\rm Proj.}}_{\hat{\imath}} & = & U_{\rm Proj.} \, \Lvec^{(0)}_i \, U^{-1}_{\rm Proj.} \ = \
\Lvec^{(0)}_{\hat{\imath}} + \Lvec^{\rm (G), Proj.}_{\hat{\imath}} \, ,
\label{L0-proj-i}
\ee
where
\be
\Lvec^{\rm (G), Proj.}_{\hat{\imath}} & \approx & 
\varepsilon_{0ijk} \lt[\tilde{R}^j{}_\al \, \xvec^\al \, \pvec^k - \xvec^j \, \tilde{R}^k{}_\al \, \pvec^\al\rt] \, ,
\label{LG-proj-i}
\ee
to leading-order in $\tilde{R}^\mu{}_\nu$,
such that the spectra of eigenvalues for $\lt\{ \lt|l, m \rt\rangle \rt\}$ and
$\lt\{ U_{\rm Proj.} \, \lt|l, m \rt\rangle \rt\}$
are identical \cite{Sakurai}.
This obvious but important property of unitarily equivalent kets in Hilbert space implies that the combination of terms
that define (\ref{LG-proj-i}) conspire to yield exactly no net changes in the observables of angular momentum
under this restriction.
As is shown below, however, a very different outcome occurs for orbital angular momentum operators defined explicitly
for a local Lorentz frame, such that the gravitational contribution generates a non-trivial interaction that needs to be
taken into account.

With this goal in mind, consider the rotation operator ${\cal D}_{\hat{\imath}}(\d \ph)$ suitably constructed to act directly on
$|\Xvec^{\hat{\mu}} (\tau, \xvec) \rangle_{\rm G}$.
For the specific case of rotation about the $z$-direction as defined by the local Lorentz frame,
\be
{\cal D}_{\hat{z}} (\d \phi) \, | \Xvec^{\hat{\mu}} \rangle & = &
\lt(1 - {i \over \hb} \, \d \ph \, \Lvec_{\hat{z}} \rt) \lt| (T, \Xvec, \Yvec, \Zvec) \rt\rangle
\nn
& = & \lt| (T, \Xvec - \Yvec \, \d \ph, \Yvec + \Xvec \, \d \ph, \Zvec) \rt\rangle \, ,
\label{z-rotation-local-Lorentz-frame}
\ee
resulting in a frame-based orbital angular momentum operator
\be
\Lvec_{\hat{\imath}} & = & \varepsilon_{0ijk} \, \Xvec^{\hat{\jmath}} \, \Pvec^{\hat{k}} \ = \
\Lvec^{(0)}_{\hat{\imath}} + \Lvec^{\rm (G)}_{\hat{\imath}} \, ,
\label{L-i}
\ee
where to leading-order in $\tilde{R}^\mu{}_\nu$,
\be
\Lvec^{\rm (G)}_{\hat{\imath}} & \approx & 
\varepsilon_{0ijk} \lt[\tilde{R}^j{}_\al \, \xvec^\al \, \pvec^k + \xvec^j \, \tilde{R}^k{}_\al \, \pvec^\al\rt]
\ \neq \ \Lvec^{\rm (G), Proj.}_{\hat{\imath}} \, .
\label{LG-i}
\ee
It becomes evident that, within the context of a local Lorentz frame, it is reasonable to expect that
$\lt| L, M \rt\rangle$ denoted by integers $(L, M)$ are eigenstates of $\Lvec^2$ and $\Lvec_{\hat{z}}$,
such that
\numparts
\be
\Lvec^2 \, \lt| L, M \rt\rangle & = & L(L + 1) \, \hb^2 \, \lt| L, M \rt\rangle \, ,
\label{L^2|LM>}
\nl
\Lvec_{\hat{z}} \, \lt| L, M \rt\rangle & = & M \, \hb \, \lt| L, M \rt\rangle \, , \qquad -L \ \leq \ M \ \leq \ L \, .
\label{Lz|LM>}
\ee
\endnumparts
However, the fact that $\Lvec^{\rm (G)}_{\hat{\imath}} \neq \Lvec^{\rm (G), Proj.}_{\hat{\imath}}$ via (\ref{LG-i})
indicates that $\lt| L, M \rt\rangle$ must be describable in general as
\be
\lt| L, M \rt\rangle & = & a \, U_{\rm Proj.} \lt| l, m \rt\rangle + b \lt| \lm \rt\rangle \, ,
\qquad \lt\langle \lm \rt| U_{\rm Proj.} \lt| l, m \rt\rangle \ = \ 0 \, ,
\label{|LM>=}
\ee
which is clearly not unitarily equivalent to $\lt| l, m \rt\rangle$, and corresponds to a physically distinct state from
one that is invisible to gravitational interactions due to local rotation.
Therefore, the local Lorentz frame rotation operators are such that
\be
{\cal D}_{\hat{\imath}} (\d \phi) & \neq & U_{\rm Proj.} \lt[{\cal D}^{(0)}_i (\d \phi)\rt] U^{-1}_{\rm Proj.} \, ,
\label{rotation-generator-local-Lorentz-frame}
\ee
implying the existence of a non-trivial local gravitational interaction to distinguish between
$\lt\{\lt| l, m \rt\rangle\rt\}$
and $\lt\{\lt| L, M \rt\rangle\rt\}$, with potential modifications in the observables of orbital angular momentum.

\section{Formalism for Orbital Angular Momentum}
\label{Formalism}

\subsection{Commutation Relations}

Having established the physical motivation to justify examination of curvature effects on orbital angular momentum,
it is necessary to now present the computational details to explicitly demonstrate their existence.
Referring now exclusively to objects defined with respect to a local Lorentz frame, for notational simplicity all indices
are now expressed in unhatted subscript form, and that $\varepsilon_{0ijk} \equiv \epsilon_{ijk}$.

It is well-known that the properties for $\Lvec_i \, (i = x, y, z)$ and the ladder operators
$\Lvec_\pm \equiv \Lvec_x \pm i \, \Lvec_y$ acting on states described by $\lt\{\lt| L, M \rt\rangle\rt\}$ must satisfy
\cite{Sakurai}
\numparts
\be
\lt[\Lvec_i \, , \Lvec_j\rt] & = & i \hb \, \epsilon_{ijk} \, \Lvec_k \, ,
\label{[Li,Lj]=ih-e(ijk)-Lk}
\nl
\lt[\Lvec_z \, , \Lvec_\pm\rt] & = & \pm \hb \, \Lvec_\pm \, ,
\label{[Lz,L+/-]=+/-h-Lz}
\nl
\lt[\Lvec_+ \, , \Lvec_-\rt] & = & 2 \, \hb \, \Lvec_z \, ,
\label{[L+,L-]=2h-Lz}
\nl
\lt[\Lvec^2 \, , \Lvec_k\rt] & = & \vec{0} \, ,
\label{[L^2,Lk]=0}
\nl
\Lvec_\pm \, \lt| L, M \rt\rangle & = & C^{\pm}_{L,M} \, \lt| L, M \pm 1 \rt\rangle \, ,
\label{L+/-|LM>}
\ee
\endnumparts
with
\be
C^{\pm}_{L,M} & = & \sqrt{(L \mp M)(L \pm M + 1)} \, .
\label{C+/-(LM)}
\ee
Simultaneously, the orbital angular momentum expressed in terms of $\Lvec^{(0)}_i \, (i = x, y, z)$ and its
corresponding ladder operators
$\Lvec^{(0)}_\pm \equiv \Lvec^{(0)}_x \pm i \, \Lvec^{(0)}_y$ acting on $\lt\{\lt| l, m \rt\rangle\rt\}$
must also satisfy
\numparts
\be
\lt[\Lvec^{(0)}_i \, , \Lvec^{(0)}_j\rt] & = & i \hb \, \epsilon_{ijk} \, \Lvec^{(0)}_k \, ,
\label{[L0i,L0j]=ih-e(ijk)-L0k}
\nl
\lt[\Lvec^{(0)}_z \, , \Lvec^{(0)}_\pm\rt] & = & \pm \hb \, \Lvec^{(0)}_\pm \, ,
\label{[L0z,L0+/-]=+/-h-L0z}
\nl
\lt[\Lvec^{(0)}_+ \, , \Lvec^{(0)}_-\rt] & = & 2 \, \hb \, \Lvec^{(0)}_z \, ,
\label{[L0+,L0-]=2h-L0z}
\nl
\lt[\Lvec_{(0)}^2 \, , \Lvec^{(0)}_k\rt] & = & \vec{0} \, ,
\label{[L0^2,L0k]=0}
\nl
\Lvec^{(0)}_\pm \, \lt| l, m \rt\rangle & = & c^{(0)\pm}_{l,m} \, \lt| l, m \pm 1 \rt\rangle \, ,
\label{L0+/-|lm>}
\ee
\endnumparts
with
\be
c^{(0)\pm}_{l,m} & = & \sqrt{(l \mp m)(l \pm m + 1)} \, .
\label{c0+/-(lm)}
\ee
It should be noted that, while $L$ must equal $l$ to conserve the orbital quantum number,
it proves useful to use these separate labels to notationally distinguish between $\lt\{\lt| l, m \rt\rangle\rt\}$
and $\lt\{\lt| L, M \rt\rangle\rt\}$, for reasons that become clear later in this paper.
Following from (\ref{L-i}), substitutions of $\Lvec_i = \Lvec^{(0)}_i + \Lvec^{\rm (G)}_i$
into (\ref{[Li,Lj]=ih-e(ijk)-Lk})--(\ref{[L+,L-]=2h-Lz}) reveal that, to first-order in $\Lvec^{\rm (G)}_i$,
\numparts
\be
\lt[\Lvec^{\rm (G)}_i \, , \Lvec^{(0)}_j\rt] + \lt[\Lvec^{(0)}_i \, , \Lvec^{\rm (G)}_j\rt]
& = & i \hb \, \epsilon_{ijk} \, \Lvec^{\rm (G)}_k \, ,
\label{[LGi,L0j]+[L0i,LGj]=ih-e(ijk)-LGk}
\nl
\lt[\Lvec^{(0)}_z \, , \Lvec^{\rm (G)}_\pm\rt] + \lt[\Lvec^{\rm (G)}_z \, , \Lvec^{(0)}_\pm\rt] & = &
\pm \, \hb \, \Lvec^{\rm (G)}_\pm \, ,
\label{[L0z,LG+/-]+[LGz,L0+/-]=(+/-)h-LGz}
\nl
\lt[\Lvec^{(0)}_+ \, , \Lvec^{\rm (G)}_-\rt] + \lt[\Lvec^{\rm (G)}_+ \, , \Lvec^{(0)}_-\rt] & = &
2 \, \hb \, \Lvec^{\rm (G)}_z \, .
\label{[L0+,LG-]+[LG+,L0-]=2h-LGz}
\ee
\endnumparts

A solution to (\ref{[LGi,L0j]+[L0i,LGj]=ih-e(ijk)-LGk}) exists, in the form
\be
\lt[\Lvec^{\rm (G)}_i \, , \Lvec^{(0)}_j\rt] & = & {i \hb \over 2} \, \epsilon_{ijk} \, \Lvec^{\rm (G)}_k \, ,
\label{[LGi,L0j]=ih/2-e(ijk)-LGk}
\ee
subject to the conditions that
\numparts
\be
\lt[\Lvec^{\rm (G)}_\pm \, , \Lvec^{(0)}_\pm\rt] & = & \vec{0} \, ,
\label{[LG+/-,L0+/-]=0}
\nl
\lt[\Lvec^{\rm (G)}_z \, , \Lvec^{(0)}_z\rt] & = & \vec{0} \, ,
\label{[LGz,L0z]=0}
\ee
\endnumparts
which need to be imposed.
For the conditions to ensure that (\ref{[LG+/-,L0+/-]=0}) and (\ref{[LGz,L0z]=0}) are
satisfied---the details of which are investigated later in this paper, it not only follows from
(\ref{[LGi,L0j]=ih/2-e(ijk)-LGk}) that 
(\ref{[L^2,Lk]=0}) is satisfied to first-order in $\Lvec^{\rm (G)}_i$,
explicit computations involving (\ref{[LGi,L0j]=ih/2-e(ijk)-LGk}), (\ref{[LG+/-,L0+/-]=0}), and (\ref{[LGz,L0z]=0})
reveal independently that
\numparts
\be
\lt[\Lvec^{(0)}_z \, , \Lvec^{\rm (G)}_\pm\rt] & = & \lt[\Lvec^{\rm (G)}_z \, , \Lvec^{(0)}_\pm\rt] \ = \
\pm \, {\hb \over 2} \, \Lvec^{\rm (G)}_\pm \, ,
\label{[L0z,LG+/-]=[LGz,L0+/-]=(+/-)h/2-LGz}
\nl
\lt[\Lvec^{\rm (G)}_\pm \, , \Lvec^{(0)}_\mp\rt] & = & \pm \, \hb \, \Lvec^{\rm (G)}_z \, ,
\label{[LG+/-,L0-/+]=h-LGz}
\ee
\endnumparts
automatically satisfying (\ref{[L0z,LG+/-]+[LGz,L0+/-]=(+/-)h-LGz}) and (\ref{[L0+,LG-]+[LG+,L0-]=2h-LGz}), respectively.

\subsection{Physical Consequences}

Some significant implications follow from (\ref{[LGi,L0j]=ih/2-e(ijk)-LGk})--(\ref{[LG+/-,L0-/+]=h-LGz})
as presented above.
To begin, consider the first commutation relation of (\ref{[L0z,LG+/-]=[LGz,L0+/-]=(+/-)h/2-LGz}) acting on
$\lt|l, m\rt\rangle$.
It is shown that
\be
\Lvec^{(0)}_z  \lt[\Lvec^{\rm (G)}_\pm \lt|l, m \rt\rangle\rt] & = &
\lt(\Lvec^{\rm (G)}_\pm \, \Lvec^{(0)}_z + \lt[\Lvec^{(0)}_z \, , \Lvec^{\rm (G)}_\pm\rt] \rt) \lt|l, m \rt\rangle
\nn
& = & \lt(m \pm {1 \over 2}\rt) \hb \lt[\Lvec^{\rm (G)}_\pm \lt|l, m \rt\rangle\rt] \, ,
\label{half-integer-z}
\ee
revealing that $\Lvec^{\rm (G)}_\pm$ is indeed a ladder operator like $\Lvec^{(0)}_\pm$, but one that raises and
lowers $m$ by {\em half-integer} steps, such that
\be
\Lvec^{\rm (G)}_\pm \lt|l, m \rt\rangle & = & c^{\rm (G) \pm}_{l,m} \lt|l, m \pm 1/2 \rt\rangle \, ,
\label{LG+/-}
\ee
with the coefficients $c^{\rm (G) \pm}_{l,m}$ to be determined.
This is very interesting because (\ref{LG+/-}) suggests that, in principle, $\Lvec^{\rm (G)}_i$ is sensitive
to half-integer $m$ projections of $\lt\{\lt|l, m \rt\rangle\rt\}$ that are otherwise considered either nonexistent
or irrelevant to orbital angular momentum.
In particular, this observation is the basis for suggesting a possible connection with the existence
of a spin-1/2 internal structure within space-time \cite{Friedman,Samuel,Williams-1}
Despite this unexpected feature, there is no a priori reason to discount its validity, given that any prediction of
a gravitational interaction in orbital angular momentum due to $\Lvec^{\rm (G)}_i$ must be very small in comparison with
a purely flat space-time computation involving $\Lvec^{(0)}_i$ alone.
A more quantitative statement to this effect follows later in this paper.

With (\ref{LG+/-}) in hand, it is also possible to determine $\Lvec^{\rm (G)}_z$ acting on $\lt|l, m\rt\rangle$,
in terms of (\ref{[L0+,LG-]+[LG+,L0-]=2h-LGz}) and (\ref{[LG+/-,L0-/+]=h-LGz}), with the result that
\be
\Lvec^{\rm (G)}_z \lt|l, m \rt\rangle & = & \lt({\cal L}^{\rm (G)}_+ + {\cal L}^{\rm (G)}_-\rt) \lt|l, m \rt\rangle
\nn
& = & \lt({\cal M}^{\rm (G)+}_{l,m} \, \hb\rt) \lt|l, m + 1/2 \rt\rangle +  \lt({\cal M}^{\rm (G)-}_{l,m} \, \hb\rt)
\, \lt|l, m - 1/2 \rt\rangle \, ,
\label{LGz-a}
\ee
%
%
where
\be
{\cal M}^{\rm (G)\pm}_{l,m} & = & \pm {1 \over 2 \, \hb^2} \lt(c^{(0)\pm}_{l,m \mp {1 \over 2}} \, c^{\rm (G)\mp}_{l,m}
- c^{(0)\pm}_{l,m} \, c^{\rm (G)\mp}_{l,m \pm 1}\rt) \, .
\label{MG}
\ee
Clearly, $\Lvec^{\rm (G)}_z$ does not generate $m$-projection eigenvalues associated with
$\lt\{\lt|l, m \rt\rangle\rt\}$, but rather shifts the $\lt|l, m \rt\rangle$ state by a half-integer above and/or
below $m$.
Following from (\ref{LGz-a}), a necessary constraint for (\ref{MG}) is that
\be
{\cal M}^{\rm (G)\pm}_{l,\pm l} & = & 0 \, ,
\label{MG-constraint}
\ee
to ensure that $\Lvec^{\rm (G)}_z \lt|l, m \rt\rangle$ remains confined to $-l \leq m \leq l$, such that
\be
\Lvec^{\rm (G)}_z \lt|l, \pm l \rt\rangle & = &  
\lt({\cal M}^{\rm (G)\mp}_{l,\pm l} \, \hb\rt) \lt|l, \pm \lt(l - 1/2\rt)\rt\rangle \, .
\label{LGz-b}
\ee
Given that $c^{(0)\pm}_{l,\pm l} = 0$ from (\ref{c0+/-(lm)}), this implies that
\be
c^{\rm (G)\mp}_{l,\pm l} & = & 0
\label{cG-constraint}
\ee
is required from (\ref{MG}).
While it is not obvious at present whether this constraint holds true, it is shown later in this paper
that (\ref{cG-constraint}) is indeed satisfied.

\section{Orbital Angular Momentum in the Presence of Local Space-Time Curvature}
\label{Orbital}

It should be clear that, while the operators $\Lvec^{\rm (G)}_\pm$ and $\Lvec^{\rm (G)}_z$ are describable in terms of
$\lt\{\lt|l, m \rt\rangle\rt\}$, they also need to be expressed within the context of $\lt\{\lt|L, M \rt\rangle\rt\}$,
since they contribute to defining the total orbital angular momentum $\Lvec_i$ that act on $\lt|L , M \rt\rangle$,
as outlined in (\ref{L^2|LM>})--(\ref{Lz|LM>}) and (\ref{[Li,Lj]=ih-e(ijk)-Lk})--(\ref{L+/-|LM>}) .
This is necessary in order to determine the coefficients $c^{\rm (G)\pm}_{l,m}$ with respect to known quantities,
and also establish the conditions in which (\ref{[LG+/-,L0+/-]=0})--(\ref{[LGz,L0z]=0}) are satisfied.
It is also necessary to determine the scalar products $\lt\langle l, m \, | L, M \rt\rangle$ relating
$\lt\{\lt|l, m \rt\rangle\rt\}$ with $\lt\{\lt|L, M \rt\rangle\rt\}$, now allowing for $m$ to take on half-integer
values, such that they can become parameters to fit with known measurement bounds.

\subsection{Constraint Equations}

To begin, recall that $\Lvec^{\rm (G)}_\pm = \Lvec_\pm - \Lvec^{(0)}_\pm$ and $\Lvec^{\rm (G)}_z = \Lvec_z -
\Lvec^{(0)}_z$, where
\numparts
\be
\Lvec_\pm & = & \sum_{M = -L}^L \lt| L, M \pm 1 \rt\rangle \, C^\pm_{L,M} \, \lt\langle L, M \rt| \, ,
\label{L+/-=}
\nl
\Lvec^{(0)}_\pm & = & \sum_{m = -l}^l \lt| l, m \pm 1 \rt\rangle \, c^{(0)\pm}_{l,m} \, \lt\langle l, m \rt| \, ,
\label{L0+/-=}
\nl
\Lvec_z & = & \sum_{M = -L}^L \lt| L, M \rt\rangle \, M \, \hb \, \lt\langle L, M \rt| \, ,
\label{Lz=}
\nl
\Lvec^{(0)}_z & = & \sum_{m = -l}^l \lt| l, m \rt\rangle \, m \, \hb \, \lt\langle l, m \rt| \, ,
\label{L0z=}
\ee
\endnumparts
keeping in mind that summations of $m$ from $-l$ to $l$ now go by half-integer steps, amounting to
$4l + 1$ terms to evaluate, while $M$ remains integer-valued with $2L + 1$ terms in the sum, and that $L = l$
to conserve the orbital quantum number.
As well, since it is true that
\be
\Lvec^{\rm (G)}_\pm & = & \sum_{m = -l}^l \lt| l, m \pm 1/2 \rt\rangle \, c^{\rm (G)\pm}_{l,m} \,
\lt\langle l, m \rt| \, ,
\label{LG+/-=}
\ee
the combination of (\ref{L+/-=}), (\ref{L0+/-=}), and (\ref{LG+/-=}) to form
$\lt\langle l, m_1 \rt| \Lvec^{\rm (G)}_\pm \lt| l, m_2 \rt\rangle$ leads to
\be
c^{\rm (G)\pm}_{l,m_2} \, \dl_{m_1, m_2 \pm {1 \over 2}} 
& = & \sum_{M = -L}^L C^\pm_{L,M} \lt\langle l, m_2 \, | L, M \rt\rangle^* \lt\langle l, m_1 \, | L, M \pm 1 \rt\rangle
\nn
&  &{}- c^{(0)\pm}_{l,m_2} 
\, \dl_{m_1, m_2 \pm 1} \, ,
\label{C1}
\ee
which for $m_1 \neq m_2 \pm {1 \over 2}$ results in a constraint equation
\be
\sum_{M = -L}^L C^\pm_{L,M} \lt\langle l, m_2 \, | L, M \rt\rangle^* \lt\langle l, m_1 \, | L, M \pm 1 \rt\rangle
- c^{(0)\pm}_{l,m_2} \, \dl_{m_1, m_2 \pm 1} & = & 0 \,
\label{C2}
\ee
for $\lt\langle l, m \, | L, M \rt\rangle$, while it follows that
\be
c^{\rm (G)\pm}_{l,m} & = & \sum_{M = -L}^L C^\pm_{L,M} \lt\langle l, m \, | L, M \rt\rangle^*
\lt\langle l, m \pm 1/2 \, | L, M \pm 1 \rt\rangle \, ,
\label{cG+/-=}
\ee
for $m_1 = m_2 \pm {1 \over 2}$.
As expected, (\ref{cG+/-=}) requires nonzero scalar products with half-integer values of $m$, in order for
$c^{\rm (G)\pm}_{l,m} \neq 0$.
In addition, the magnitude of $c^{\rm (G)\pm}_{l,m}$ can be determined according to
$\lt|c^{\rm (G)\pm}_{l,m}\rt|^2 = \lt\langle l, m \rt| \Lvec^{\rm (G)}_\mp \, \Lvec^{\rm (G)}_\pm \lt| l, m \rt\rangle$,
leading to 
\be
\lt|c^{\rm (G)\pm}_{l,m}\rt|^2 & = & \lt|c^{(0)\pm}_{l,m}\rt|^2
+ \sum_{M = -L}^L \lt\{\lt|C^\pm_{L,M}\rt|^2 \lt|\lt\langle l, m \, | L, M \rt\rangle \rt|^2 \rt.
\nn
&  &{} - 2 \lt. {\rm Re} \lt[C^\pm_{L,M} \, c^{(0)\pm}_{l,m} \, \lt\langle l, m \pm 1 \, | L, M \pm 1 \rt\rangle^*
\lt\langle l, m \, | L, M \rt\rangle \rt] \rt\} \, .
\label{|cG+/-|^2=}
\ee

Considering now the commutators (\ref{[LG+/-,L0+/-]=0})--(\ref{[LGz,L0z]=0}) that are assumed true when used in
conjunction with (\ref{[LGi,L0j]=ih/2-e(ijk)-LGk}), it is a straightforward matter to evaluate them directly using
(\ref{L+/-=})--(\ref{L0z=}).
Starting with (\ref{[LG+/-,L0+/-]=0}), it follows that
\be
\lt[\Lvec^{\rm (G)}_\pm \, , \Lvec^{(0)}_\pm\rt] & = & \sum_{m = -l}^l
\lt[c^{(0)\pm}_{l,m} \, c^{\rm (G)\pm}_{l,m \pm 1} - c^{(0)\pm}_{l,m \pm {1 \over 2}} \, c^{\rm (G)\pm}_{l,m} \rt]
\lt| l, m \pm 3/2 \rt\rangle \lt\langle l, m \rt|
\nn
& \stackrel{?}{=} & \vec{0} \, ,
\label{[LG+/-,L0+/-]=0-evaluation}
\ee
which is always true if the prefactor for each $m$ is set to zero.
Therefore, (\ref{[LG+/-,L0+/-]=0-evaluation}) is satisfied at the operator level, provided that the constraint equation
\be
\sum_{M = -L}^L & C^\pm_{L,M} &
\lt[c^{(0)\pm}_{l,m} \, \lt\langle l, m \pm 1 \, | L, M \rt\rangle^* \lt\langle l, m \pm 3/2 \,
| L, M \pm 1 \rt\rangle \rt.
\nn
&  &{} - \lt. c^{(0)\pm}_{l,m \pm {1 \over 2}} \, \lt\langle l, m \, | L, M \rt\rangle^*
\lt\langle l, m \pm 1/2 \, | L, M \pm 1 \rt\rangle \rt] \ = \ 0
\label{C3}
\ee
for $\lt\langle l, m \, | L, M \rt\rangle$ is satisfied, using (\ref{cG+/-=}) in place of $c^{\rm (G)\pm}_{l,m}$.

As for (\ref{[LGz,L0z]=0}), a similar approach with the incorporation of
\be
\lt|L, M \rt\rangle & = & \sum_{m = -l}^l \lt\langle l, m \, | L, M \rt\rangle \lt|l, m \rt\rangle
\label{|LM>=sum(|lm>)}
\ee
results in
\be
\lt[\Lvec^{\rm (G)}_z \, , \Lvec^{(0)}_z\rt] & = & \sum_{m = -l}^l \sum_{m' = -l}^l
\lt\{\hb^2 \sum_{M = -L}^L M (m' - m) \lt\langle l, m' \, | L, M \rt\rangle^* \lt\langle l, m \, | L, M \rt\rangle \rt\}
\nn
&  &{} \times
\lt|l, m \rt\rangle \lt\langle l, m' \rt| \ \stackrel{?}{=} \ \vec{0} \, .
\label{[LGz,L0z]=0-evaluation}
\ee

\subsection{Evaluation of the Constraint Equations}

Unlike (\ref{[LG+/-,L0+/-]=0-evaluation}), it is not obvious at present whether (\ref{[LGz,L0z]=0-evaluation})
holds true.
In part to address this issue, it is very useful to introduce a physically motivated simplification by
letting (\ref{|LM>=sum(|lm>)}) approximate to
\be
\lt|L, M \rt\rangle & \approx & \al_M \lt|l, M \rt\rangle + \bt^{(+)}_M \lt|l, M + 1/2 \rt\rangle
+ \bt^{(-)}_M \lt|l, M - 1/2 \rt\rangle \, ,
\label{|LM>=sum(|lm>)-approx}
\ee
or equivalently
\be
\lt|l, m \rt\rangle & \approx & \al^*_m \lt|L, m \rt\rangle + \bt^{(+)*}_{m - {1 \over 2}} \lt|L, m - 1/2 \rt\rangle
+ \bt^{(-)*}_{m + {1 \over 2}} \lt|L, m + 1/2 \rt\rangle \, ,
\label{|lm>=sum(|LM>)-approx}
\ee
with
\be
\lt\langle L, M \, | L, M \rt\rangle & = &
\lt|\al_M\rt|^2 + \lt|\bt^{(+)}_M\rt|^2 + \lt|\bt^{(-)}_M\rt|^2 \ \approx \ 1 \,
\label{<LM|LM>=1}
\ee
such that $ \lt|\bt^{(\pm)}_M\rt|^2 \ll \lt|\al_M\rt|^2$ for each integer-valued $M$, and $\bt^{(\pm)}_{\pm L} = 0$.

With (\ref{|LM>=sum(|lm>)-approx}), it is possible to conveniently express (\ref{L0z=}) in the form
\be
\Lvec^{(0)}_z & = & \hb \sum_{M = -L}^L \lt|L, M \rt\rangle
\lt[M \, \lt|\al_M\rt|^2 + \lt(M + {1 \over 2}\rt) \lt|\bt^{(+)}_M\rt|^2 \rt.
\nn
&  &{} \hspace{4.4cm} \lt. + \lt(M - {1 \over 2}\rt) \lt|\bt^{(-)}_M\rt|^2\rt] \lt\langle L, M \rt|
\nn
&  &{} + \hb \sum_{M = -L}^L \lt|L, M \rt\rangle \lt(M + {1 \over 2}\rt) \bt^{(+)*}_M \, \bt^{(-)}_{M + 1}
\, \lt\langle L, M + 1 \rt|
\nn
&  &{} + \hb \sum_{M = -L}^L \lt|L, M \rt\rangle \lt(M - {1 \over 2}\rt) \bt^{(-)*}_M \, \bt^{(+)}_{M - 1}
\, \lt\langle L, M - 1 \rt| \, .
\label{L0z/-(M-basis)=}
\ee
Therefore, it becomes straightforward to show that
\be
\lt[\Lvec^{\rm (G)}_z \, , \Lvec^{(0)}_z\rt] & = & \hb^2 \sum_{M = -L}^L \lt|L, M \rt\rangle
\lt[\lt(M - {1 \over 2}\rt) \, \bt^{(-)*}_M \, \bt^{(+)}_{M - 1} \, \lt\langle L, M - 1\rt| \rt.
\nn
&  &{} \hspace{2.4cm}
- \lt. \lt(M + {1 \over 2}\rt) \, \bt^{(+)*}_M \, \bt^{(-)}_{M + 1} \, \lt\langle L, M + 1\rt| \rt]
\, .
\label{[LGz,L0z]=}
\ee
Clearly, (\ref{[LGz,L0z]=}) is nonzero, with a structure that is somewhat analogous to
angular momentum orthogonal to the $z$-direction \cite{Bransden}.
However, (\ref{[LGz,L0z]=}) is only a {\em second-order} operator in $\bt^{(\pm)}_M$,
 while the right-hand side of (\ref{[LGi,L0j]=ih/2-e(ijk)-LGk})
is necessarily a {\em first-order} operator.
Therefore, (\ref{[LGz,L0z]=}) satisfies (\ref{[LGz,L0z]=0}) within this approximation structure
and still satisfies (\ref{[LGi,L0j]+[L0i,LGj]=ih-e(ijk)-LGk}) outright.
In retrospect, observing this discrepancy as it concerns (\ref{[LGz,L0z]=0}) is not surprising,
considering that space-time curvature has the effect of introducing other breakdowns in
quantum mechanical concepts, such as the non-Hermiticity of the Dirac Hamiltonian for spin-1/2
particles in a gravitational background \cite{Singh-Mobed-zitt,Parker}.

Turning attention now to (\ref{C3}), the constraint equation for
$\lt[\Lvec^{\rm (G)}_\pm \, , \Lvec^{(0)}_\pm\rt] = \vec{0}$, substitution of
(\ref{|LM>=sum(|lm>)-approx}) results in
\be
c^{(0)\pm}_{l,m} \lt[C^\pm_{L,m \pm 1} \, \al^*_{m \pm 1}
\lt(\bt^{(+)}_{m \pm {3 \over 2} - {1 \over 2}} \, \dl_{m \pm {3 \over 2} - {1 \over 2}, m \pm 2}
+ \bt^{(-)}_{m \pm {3 \over 2} + {1 \over 2}} \, \dl_{m \pm {3 \over 2} + {1 \over 2}, m \pm 2} \rt) \rt.
\nn
\hspace{0.6cm}
{}+ C^\pm_{L,m \pm 1 - {1 \over 2}} \, \bt^{(+)*}_{m \pm 1 - {1 \over 2}} \, \al_{m \pm {3 \over 2}} \,
\dl_{m \pm {3 \over 2}, m \pm 2 - {1 \over 2}}
\nn
\hspace{0.55cm} \lt.
{}+ C^\pm_{L,m \pm 1 + {1 \over 2}} \, \bt^{(-)*}_{m \pm 1 + {1 \over 2}} \, \al_{m \pm {3 \over 2}} \,
\dl_{m \pm {3 \over 2}, m \pm 2 + {1 \over 2}}
\rt]
\nn \nn
{} - c^{(0)\pm}_{l,m \pm {1 \over 2}} \lt[C^\pm_{L,m} \, \al^*_m
\lt(\bt^{(+)}_{m \pm {1 \over 2} - {1 \over 2}} \, \dl_{m \pm {1 \over 2} - {1 \over 2}, m \pm 1}
+ \bt^{(-)}_{m \pm {1 \over 2} + {1 \over 2}} \, \dl_{m \pm {1 \over 2} + {1 \over 2}, m \pm 1} \rt) \rt.
\nn
\hspace{1.35cm}
{}+ C^\pm_{L,m - {1 \over 2}} \, \bt^{(+)*}_{m - {1 \over 2}} \, \al_{m \pm {1 \over 2}} \,
\dl_{m \pm {1 \over 2}, m \pm 1 - {1 \over 2}}
\nn
\hspace{1.30cm} \lt.
{}+ C^\pm_{L,m + {1 \over 2}} \, \bt^{(-)*}_{m + {1 \over 2}} \, \al_{m \pm {1 \over 2}} \,
\dl_{m \pm {1 \over 2}, m \pm 1 + {1 \over 2}}
\rt] \ = \ 0 \, .
\label{C3-1}
\ee
When evaluated for the upper and lower signs separately, (\ref{C3-1}) can be expressed in a single constraint
equation as
\be
c^{(0)\pm}_{l,m} \lt[C^\pm_{L,m \pm 1} \, \al^*_{m \pm 1} \, \bt^{(\mp)}_{m \pm 2}
+ C^\pm_{L,m \pm {1 \over 2}} \, \al_{m \pm {3 \over 2}} \, \bt^{(\pm)*}_{m \pm {1 \over 2}}\rt]
\nn \nn
- c^{(0)\pm}_{l,m \pm {1 \over 2}} \lt[C^\pm_{L,m} \, \al^*_m \, \bt^{(\mp)}_{m \pm 1}
+ C^\pm_{L,m \mp {1 \over 2}} \, \al_{m \pm {1 \over 2}} \, \bt^{(\pm)*}_{m \mp {1 \over 2}}\rt] \ = \ 0 \, .
\label{C3-1a}
\ee
The structure of (\ref{C3-1a}) demands that, for $m$ equal to an integer,
\be
c^{(0)\pm}_{l,m} \, C^\pm_{L,m \pm 1} \, \al^*_{m \pm 1} \, \bt^{(\mp)}_{m \pm 2}
- c^{(0)\pm}_{l,m \pm {1 \over 2}} \, C^\pm_{L,m} \, \al^*_m \, \bt^{(\mp)}_{m \pm 1} \ = \ 0 \, ,
\label{C3-1a-integer}
\ee
while for $m$ equal to a half-integer,
\be
c^{(0)\pm}_{l,m} \, C^\pm_{L,m \pm {1 \over 2}} \, \al_{m \pm {3 \over 2}} \, \bt^{(\pm)*}_{m \pm {1 \over 2}}
- c^{(0)\pm}_{l,m \pm {1 \over 2}} \, C^\pm_{L,m \mp {1 \over 2}} \, \al_{m \pm {1 \over 2}} \,
\bt^{(\pm)*}_{m \mp {1 \over 2}} \ = \ 0 \, .
\label{C3-1a-half-integer}
\ee

By letting $m \rightarrow m \mp {1 \over 2}$ in (\ref{C3-1a-half-integer}) for direct comparison with
(\ref{C3-1a-integer}) and recognizing that $C^\pm_{L,m} = c^{(0)\pm}_{l,m}$ for $L = l$ and $M = m$, it follows that
\be
\lt(\al_{M \pm 1} \over \al_M\rt)^* & = & \lt(C^{\pm}_{L,M \pm {1 \over 2}} \over C^{\pm}_{L,M \pm 1}\rt)
\lt(\bt^{(\mp)}_{M \pm 1} \over \bt^{(\mp)}_{M \pm 2}\rt) \ = \
\lt(C^{\pm}_{L,M \mp 1} \over C^{\pm}_{L,M \mp {1 \over 2}}\rt)
\lt(\bt^{(\pm)}_{M \mp 1} \over \bt^{(\pm)}_M\rt) \, ,
\label{C3-b}
\ee
which leads to a recursion relation for $\bt^{(\pm)}_M$, of the form
\be
\bt^{(\mp)}_{M \pm 2} & = & \lt(C^+_{L,M \pm {1 \over 2}} \over C^+_{L,M \pm 1}\rt)
\lt(C^-_{L,M \pm {1 \over 2}} \over C^-_{L,M \pm 1}\rt) \lt(\al_{M \mp 1} \over \al_{M \pm 1}\rt)^* \,
\bt^{(\mp)}_M \, ,
\qquad (\bt^{(\pm)}_{\pm L} \ = \ 0) \, .
\label{beta-recursion}
\ee
Without knowing anything further about this recursion relation, it is immediately obvious that
(\ref{beta-recursion}) is subject to the restriction $-(L - 2) \leq M \leq L - 2$, which implies a minimum allowable
value of $L = 2$ in order for any gravitational corrections of orbital angular momentum to occur within this framework.
This is also the basis for suggesting a deep connection with the properties of the graviton for
linearized general relativity, since it is anticipated to propagate with spin angular momentum $S = 2$,
matching perfectly with the constraint conditions of (\ref{beta-recursion}).

For the final constraint equation (\ref{C2}) to consider, substitution of (\ref{|LM>=sum(|lm>)-approx}) leads to
\be
\lt[C^{\pm}_{L, m_1 \mp 1} \, \al_{m_1} \, \al^*_{m_1 \mp 1} - c^{(0)\pm}_{l,m_1 \mp 1} \rt]
\dl_{m_1, m_2 \pm 1}
\nn \nn
+ \lt[C^{\pm}_{L, m_1 \mp {3 \over 2}} \, \al^*_{m_1 \mp {3 \over 2}} \, \bt^{(\pm)}_{m_1 \mp {1 \over 2}}
+ C^{\pm}_{L, m_1 \mp 1} \, \al_{m_1} \, \bt^{(\mp)*}_{m_1 \mp 1} \rt]
\dl_{m_1, m_2 \pm {3 \over 2}} \ = \ 0 \, ,
\label{C2-1}
\ee
subject to $m_1 \neq m_2 \pm {1 \over 2}$.
It is evident from (\ref{C2-1}) that the only physically relevant constraint to be satisfied is the
recurrence relation corresponding to $m_1 = m_2 \pm 1$, since while the one corresponding to
$m_1 = m_2 \pm {3 \over 2}$ demands the existence of $\al_M$ and $\bt^{(\pm)}_M$ for {\em half-integer} $M$
to ensure that non-trivial {\em integer-valued} $M$ terms can appear,
such quantities are not accessed within the framework of this problem and can be ignored.
Therefore, the recursion relations for $\al_M$ are
\numparts
\be
\al_{M \pm 1} & = & {\al_M \over \lt|\al_M\rt|^2} \, ,
\label{alpha-recursion-1}
\nl
\al_{M \pm 2} & = & \al_M \, ,
\label{alpha-recursion-2}
\ee
\endnumparts
and from substituting (\ref{alpha-recursion-1}) into (\ref{C3-b}) and (\ref{beta-recursion}), the recursion relations
for $\bt^{(\pm)}_M$ are
\numparts
\be
\bt^{(\mp)}_{M \pm 1} & = & \lt(C^{\mp}_{L,M \pm {1 \over 2}} \over C^{\mp}_{L,M \pm 1}\rt)
\, {\bt^{(\mp)}_M \over \lt|\al_M\rt|^2} \, ,
\hspace{2.2cm} -(L - 2) \, \leq \, M \, \leq \, L - 2 \, ,
\label{beta-recursion-1}
\nl
\bt^{(\mp)}_{M \pm 2} & = & \lt(C^{\pm}_{L,M \pm {1 \over 2}} \over C^{\pm}_{L,M \pm 1}\rt)
\, \lt|\al_M\rt|^2 \, \bt^{(\mp)}_{M \pm 1} \,
\label{beta-recursion-2a}
\nl
& = & \lt(C^+_{L,M \pm {1 \over 2}} \over C^+_{L,M \pm 1}\rt)
\lt(C^-_{L,M \pm {1 \over 2}} \over C^-_{L,M \pm 1}\rt) \, \bt^{(\mp)}_M \, ,
\quad -(L - 2) \, \leq \, M \, \leq \, L - 2 \, ,
\label{beta-recursion-2b}
\ee
\endnumparts
with $\lt|\al_M\rt|^2 \approx 1 - \lt|\bt^{(+)}_M\rt|^2 - \lt|\bt^{(-)}_M\rt|^2$ from (\ref{<LM|LM>=1}).
With $\bt^{(\pm)}_{\pm L} = 0$ as a boundary condition, only $\bt^{(\pm)}_{\pm(L - 1)}$ is left to specify separately.
This is accomplished by first letting $M \rightarrow M \mp 1$ in (\ref{beta-recursion-2a}), such that
\be
\bt^{(\mp)}_{M \pm 1} & = & \lt(C^{\pm}_{L,M \mp {1 \over 2}} \over C^{\pm}_{L,M}\rt)
\, \lt|\al_{M \mp 1}\rt|^2 \, \bt^{(\mp)}_M \, . 
\label{beta-recursion-2c}
\ee
Therefore, by substituting $M = \mp(L - 1)$ into (\ref{beta-recursion-2c}) and re-arranging, it follows that
\be
\bt^{(\pm)}_{\pm(L - 1)} & = & \lt(C^{\mp}_{L, \pm(L - 1)} \over C^{\mp}_{L, \pm(L - {1 \over 2})}\rt)
\, {\bt^{(\pm)}_{\pm(L - 2)} \over \lt|\al_{\pm L}\rt|^2} \, .
\label{beta(+/-)-(L-1)=}
\ee
With (\ref{alpha-recursion-1})--(\ref{beta(+/-)-(L-1)=}), it becomes clear that all of the
$\al_M$ and $\bt^{(\pm)}_M$ can be computed
in relation to $\al_0$ and $\bt^{(\pm)}_0$ as input parameters to be determined from experimental data.

\subsection{Evaluation of the Gravitational Ladder Coefficients}

It remains to present the computation of the gravitational ladder coefficients $c^{\rm (G)\pm}_{l,m}$
and their magnitudes according to (\ref{cG+/-=}) and (\ref{|cG+/-|^2=}), respectively,
based on the approximation employed for $\lt| L, M \rt\rangle$ via (\ref{|LM>=sum(|lm>)-approx}).
Not surprisingly, the coefficients are defined in accordance with the choice of $m$ as either
integer- or half-integer-valued.
Therefore, with use of (\ref{alpha-recursion-1}) and (\ref{beta-recursion-1}),
it follows from a straightforward evaluation of (\ref{cG+/-=}) that
\be
c^{\rm (G)\pm}_{l,m} & = & c^{(0)\pm}_{l,m} \, \al^*_m \, \bt^{(\mp)}_{m \pm 1}
\ = \ \lt(c^{(0)\pm}_{l,m} \over \lt|\al_m\rt|^2\rt) \lt(c^{(0)\mp}_{l,m \pm {1 \over 2}} \over c^{(0)\mp}_{l,m \pm 1}\rt)
\, \al^*_m \, \bt^{(\mp)}_m
\label{cG+/-(integer)}
\ee
for $m$ an integer, while
\be
c^{\rm (G)\pm}_{l,m} & = & c^{(0)\pm}_{l,m \mp {1 \over 2}} \, \al_{m \pm {1 \over 2}} \,
\bt^{(\pm)*}_{m \mp {1 \over 2}}
\ = \ \lt(c^{(0)\pm}_{l,m \mp {1 \over 2}} \over \lt|\al_{m \mp {1 \over 2}}\rt|^2\rt) \, \al_{m \mp {1 \over 2}} \,
\bt^{(\pm)*}_{m \mp {1 \over 2}}
\label{cG+/-(half-integer)}
\ee
for $m$ a half-integer.
Clearly, both (\ref{cG+/-(integer)}) and (\ref{cG+/-(half-integer)}) vanish in the limit as
$\lt|\bt^{(\pm)}_M\rt| \rightarrow 0$, as expected.
As well, (\ref{cG+/-(integer)}) satisfies the condition that $c^{\rm (G)\mp}_{l,\pm l} = 0$ as claimed in (\ref{cG-constraint}),
such that (\ref{MG-constraint}) is also satisfied, ensuring that $\Lvec^{\rm (G)}_z \lt|l, m \rt\rangle$ is confined to
$-l \leq m \leq l$.

It is interesting to note from (\ref{alpha-recursion-1}) and (\ref{beta-recursion-1}) that each iteration of $\al_M$
and $\bt^{(\pm)}_M$ preserves the same phase angle of its predecessor, as denoted by
$\tan \gm_M = {\rm Im}(\al_M)/{\rm Re}(\al_M)$ and $\tan \dl^{(\pm)}_M =
{\rm Im}(\bt^{(\pm)}_M)/{\rm Re}(\bt^{(\pm)}_M)$, though it is still possible for $\al_M$ and $\bt^{(\pm)}_M$ to have
a relative phase difference.
As a result, both (\ref{cG+/-(integer)}) and (\ref{cG+/-(half-integer)})
demonstrate the existence of a generally non-zero phase angle associated with $c^{\rm (G)\pm}_{l,m}$,
given by
\be
\tan \vph_m
& = &{} -{\lt(\tan \gm_m - \tan \dl^{(\mp)}_m\rt) \over \lt(1 + \tan \gm_m \, \tan \dl^{(\mp)}_m\rt)}
\ = \ - \tan \lt(\gm_m - \dl^{(\mp)}_m\rt)
\label{tan-phi(integer)}
\ee
for integer-valued $m$, while
\be
\tan \vph_m
& = &{} {\lt(\tan \gm_{m \mp {1 \over 2}} - \tan \dl^{(\pm)}_{m \mp {1 \over 2}}\rt)
\over \lt(1 + \tan \gm_{m \mp {1 \over 2}} \, \tan \dl^{(\pm)}_{m \mp {1 \over 2}}\rt)}
\ = \ \tan \lt(\gm_{m \mp {1 \over 2}} - \dl^{(\pm)}_{m \mp {1 \over 2}}\rt)
\label{tan-phi(half-integer)}
\ee
for half-integer-valued $m$, with $\tan \vph_{m \pm 1} = \tan \vph_m$ for both cases.

Regarding the magnitude of the gravitational ladder coefficients in terms of (\ref{|LM>=sum(|lm>)-approx})
and (\ref{|lm>=sum(|LM>)-approx}), recall that
$\lt|c^{\rm (G)\pm}_{l,m}\rt|^2 = \lt\langle l, m \rt| \Lvec^{\rm (G)}_\mp \, \Lvec^{\rm (G)}_\pm \lt| l, m \rt\rangle$,
where
\be
\lefteqn{\Lvec^{\rm (G)}_\mp \, \Lvec^{\rm (G)}_\pm \ = \ \Lvec^{(0)}_\mp \, \Lvec^{(0)}_\pm
+ \sum_{M = -L}^L \lt|L, M \rt\rangle \lt|C^\pm_{L,M}\rt|^2 \lt\langle L, M \rt| }
\nn
&&{} -  \sum_{M = -L}^L  \lt|C^\pm_{L,M}\rt|  \lt|c^{(0)\pm}_{l,M}\rt| \lt[\lt|L, M \rt\rangle \al_{M \pm 1} \lt\langle L, M \rt|
+ \lt|L, M \rt\rangle \al^*_{M \pm 1} \lt\langle L, M \rt| \rt]
\nn
&&{} -  \sum_{M = -L}^L  \lt|C^\pm_{L,M}\rt|  \lt|c^{(0)\pm}_{l,M + {1 \over 2}}\rt|
\lt[\lt|L, M + 1/2 \rt\rangle \bt^{(+)}_{M \pm 1} \lt\langle L, M \rt| \rt.
\nn
&&{} \lt. \hspace{5.0cm} + \lt|L, M \rt\rangle \bt^{(+)*}_{M \pm 1} \lt\langle L, M + 1/2\rt| \rt]
\nn
&&{} -  \sum_{M = -L}^L  \lt|C^\pm_{L,M}\rt|  \lt|c^{(0)\pm}_{l,M - {1 \over 2}}\rt|
\lt[\lt|L, M - 1/2 \rt\rangle \bt^{(-)}_{M \pm 1} \lt\langle L, M \rt| \rt.
\nn
&&{} \lt. \hspace{5.0cm} + \lt|L, M \rt\rangle \bt^{(-)*}_{M \pm 1} \lt\langle L, M - 1/2\rt| \rt] \, ,
\label{LG(-/+)LG(+/-)=}
\ee
and
\be
\lefteqn{\Lvec^{(0)}_\mp \, \Lvec^{(0)}_\pm \ = \ \sum_{m = -l}^l \lt|l, m \rt\rangle \lt|c^{(0)\pm}_{l,m}\rt|^2 \lt\langle l, m \rt| }
\nn
&&{} = \ \sum_{M = -L}^L \lt\{ \lt|L, M \rt\rangle \lt(\lt|\al_M\rt|^2 \lt|c^{(0)\pm}_{l,M}\rt|^2
+ \lt|\bt^{(+)}_M\rt|^2 \lt|c^{(0)\pm}_{l,M + {1 \over 2}}\rt|^2
+ \lt|\bt^{(-)}_M\rt|^2 \lt|c^{(0)\pm}_{l,M - {1 \over 2}}\rt|^2\rt) \rt.
\nn
&&{} \hspace{1.5cm} \times \lt\langle L, M \rt|
+ \lt|L, M \rt\rangle \bt^{(+)*}_M \, \bt^{(-)}_{M + 1} \, \lt|c^{(0)\pm}_{l,M + {1 \over 2}}\rt|^2 \lt\langle L, M + 1 \rt|
\nn
&&{} \hspace{3.2cm} \lt. + \lt|L, M \rt\rangle \bt^{(-)*}_M \, \bt^{(+)}_{M - 1} \, \lt|c^{(0)\pm}_{l,M - {1 \over 2}}\rt|^2
\lt\langle L, M - 1 \rt| \rt\} \, .
\label{L0(-/+)L0(+/-)=}
\ee
It happens that $\lt|c^{\rm (G)\pm}_{l,m}\rt|^2$ is also dependent on whether $m$ is an integer or
half-integer.
Therefore, from (\ref{LG(-/+)LG(+/-)=}) and (\ref{L0(-/+)L0(+/-)=}), it follows that
\be
\lt|c^{\rm (G)\pm}_{l,m}\rt|^2 & = & \lt\{\lt(1 + \lt|\al_m\rt|^2 \rt)\lt|\al_m\rt|^2 - 2\rt\}  \lt|c^{(0)\pm}_{l,m}\rt|^2
\nn \nn
&  &{} + \lt\{\lt|\bt^{(+)}_m\rt|^2 \lt|c^{(0)\pm}_{l,m + {1 \over 2}}\rt|^2
+ \lt|\bt^{(-)}_m\rt|^2 \lt|c^{(0)\pm}_{l,m - {1 \over 2}}\rt|^2\rt\} \lt|\al_m\rt|^2
\label{|cG+/-|^2(integer)}
\ee
for $m$ integer, while
\be
\lt|c^{\rm (G)\pm}_{l,m}\rt|^2 & = & \lt\{\lt(1 + \lt|\al_{m - {1 \over 2}}\rt|^2\rt)\lt|c^{(0)\pm}_{l,m - {1 \over 2}}\rt|^2
+ \lt|\bt^{(-)}_{m - {1 \over 2}}\rt|^2 \lt|c^{(0)\pm}_{l,m - 1}\rt|^2 \rt.
\nn
&  &{} \lt. \hspace{2.5cm} + \lt(\lt|\bt^{(+)}_{m - {1 \over 2}}\rt|^2 + \lt|\bt^{(-)}_{m + {1 \over 2}}\rt|^2\rt)
\lt|c^{(0)\pm}_{l,m}\rt|^2 \rt\} \lt|\bt^{(+)}_{m - {1 \over 2}}\rt|^2
\nn \nn
&  &{} + \lt\{\lt(1 + \lt|\al_{m + {1 \over 2}}\rt|^2\rt)\lt|c^{(0)\pm}_{l,m + {1 \over 2}}\rt|^2
+ \lt|\bt^{(+)}_{m + {1 \over 2}}\rt|^2 \lt|c^{(0)\pm}_{l,m + 1}\rt|^2 \rt.
\nn
&  &{} \lt. \hspace{2.5cm} + \lt(\lt|\bt^{(+)}_{m - {1 \over 2}}\rt|^2 + \lt|\bt^{(-)}_{m + {1 \over 2}}\rt|^2\rt)
\lt|c^{(0)\pm}_{l,m}\rt|^2 \rt\} \lt|\bt^{(-)}_{m + {1 \over 2}}\rt|^2
\nn \nn
&  &{} - 2 \, \lt|c^{(0)\pm}_{l,m}\rt| \lt\{\lt|c^{(0)\pm}_{l,m + {1 \over 2}}\rt|  \, {\rm Re} \lt(\bt^{(-)*}_{m \pm 1 + {1 \over 2}} \,
\bt^{(-)}_{m + {1 \over 2}} \rt) \rt.
\nn
&  &{} \lt. \hspace{1.7cm} + \lt|c^{(0)\pm}_{l,m - {1 \over 2}}\rt| \, {\rm Re} \lt(\bt^{(+)*}_{m \pm 1 - {1 \over 2}} \,
\bt^{(+)}_{m - {1 \over 2}} \rt) \rt\} \, .
\label{|cG+/-|^2(half-integer)}
\ee
for half-integer $m$.
It is self-evident that $\lt|c^{\rm (G)\pm}_{l,m}\rt|^2 \rightarrow 0$ for both (\ref{|cG+/-|^2(integer)})
and (\ref{|cG+/-|^2(half-integer)}) in the limit as $\lt|\bt^{(\pm)}_M\rt| \rightarrow 0$ and $\lt|\al_M\rt| \rightarrow 1$,
again ensuring that all gravitationally induced expressions involving orbital angular momentum
vanish smoothly in the flat space-time limit.

\subsection{Physical Consequences for Observables}

It is important to understand the physical consequences for the observables of orbital angular momentum
when incorporating local space-time curvature in this proposed fashion.
This entails performing the direct computation of $\Lvec^2 \lt|L, M \rt\rangle$ and $\Lvec_z \lt|L, M \rt\rangle$
to show the dependence of $\al_M$ and $\bt^{(\pm)}_M$ on the expressed quantities.
As shown explicitly below, it follows that while both $\lt[\Lvec^2 - L(L + 1) \, \hb^2\rt] \lt|L, M \rt\rangle$ and
$\lt[\Lvec_z - M \, \hb\rt]\lt|L, M \rt\rangle$ equal zero to first-order in
$\bt^{(\pm)}_M$, the second-order terms which survive are {\em not} proportional to $\lt|L, M \rt\rangle$,
indicating that $\lt\{ \lt|L, M \rt\rangle \rt\}$ can no longer be classified as the eigenstates
of $\Lvec^2$ and $\Lvec_z$.
Given (\ref{[LGz,L0z]=}) for the evaluation of $\lt[\Lvec^{\rm (G)}_z \, , \Lvec^{(0)}_z\rt]$,
this should not come as a surprise.
Nonetheless, it is necessary to fully understand all the consequences that result from introducing local
curvature effects into the current description of orbital angular momentum.

To illustrate this more fully, recall that to first-order in $\Lvec^{\rm (G)}_i$
\be
\Lvec^2 & = & \Lvec^2_{(0)} + \lt\{\Lvec^{\rm (G)}_i, \Lvec^{(0)}_i\rt\} \, ,
\label{L^2=}
\ee
where it is shown that
\be
\lt\{\Lvec^{\rm (G)}_i, \Lvec^{(0)}_i\rt\} & = & \Lvec^{(0)}_+ \, \Lvec^{\rm (G)}_- + \Lvec^{(0)}_- \, \Lvec^{\rm (G)}_+
+ 2 \, \Lvec^{(0)}_z \, \Lvec^{\rm (G)}_z + \lt[\Lvec^{\rm (G)}_z, \Lvec^{(0)}_z\rt] \,
\label{{LGi,L0i}-formal}
\ee
and
\be
\Lvec^{\rm (G)}_z & = & \sum_{m = -l}^l \lt(
\lt| l, m + 1/2\rt\rangle \, {\cal M}^{\rm (G)+}_{l,m} \, \hb \, \lt\langle l, m \rt| \rt.
\nn
&  &{} \hspace{0.6cm} + \lt. \lt| l, m - 1/2\rt\rangle \, {\cal M}^{\rm (G)-}_{l,m} \, \hb \, \lt\langle l, m \rt| \rt) \, ,
\label{LGz=}
\ee
from (\ref{LGz-a}) and (\ref{MG}).
Knowing (\ref{L0+/-=}), (\ref{L0z=}), and (\ref{LG+/-=}) in terms of $\lt\{\lt|l, m \rt\rangle\rt\}$,
their equivalent expressions in terms of $\lt\{\lt|L, M \rt\rangle\rt\}$ in combination with (\ref{[LGz,L0z]=})
results in
\be
\lefteqn{\lt\{\Lvec^{\rm (G)}_i, \Lvec^{(0)}_i\rt\} \ = \ \sum_{M = -L}^L
\lt|L, M \rt\rangle \lt\{
\al_M \lt(\bt^{(+)*}_M \, {\cal F}^{\rm (G)+}_{L,M} + \bt^{(-)*}_M \, {\cal F}^{\rm (G)-}_{L,M}\rt) \hb^2 \rt\}
\lt\langle L, M \rt| }
\nn
&&{} + \sum_{M = -L}^L
\lt|L, M \rt\rangle \lt\{ \bt^{(-)*}_M \lt[\al_{M-1} \, {\cal F}^{\rm (G)+}_{L,M-1}
+ \lt(M - {1 \over 2}\rt) \, \bt^{(+)}_{M-1}\rt] \hb^2 \rt\} \lt\langle L, M - 1\rt|
\nn
&&{} + \sum_{M = -L}^L
\lt|L, M \rt\rangle \lt\{ \bt^{(+)*}_M \lt[\al_{M+1} \, {\cal F}^{\rm (G)-}_{L,M+1}
- \lt(M + {1 \over 2}\rt) \, \bt^{(-)}_{M+1}\rt] \hb^2 \rt\} \lt\langle L, M + 1\rt| \, ,
\nn
\label{{LGi,L0i}=}
\ee
where
\be
{\cal F}^{\rm (G)\pm}_{l,m} & \equiv & (2 \, m \pm 1) \, {\cal M}^{\rm (G)\pm}_{l,m} +
{1 \over \hb^2} \, c^{(0)\pm}_{l,m \mp {1 \over 2}} \, c^{\rm (G)\pm}_{l,m}
 \,
\label{F=}
\ee
is first-order in $\bt^{(\pm)}_M$.
Therefore, it follows that
\be
\Lvec^2 \lt|L, M \rt\rangle & = & \lt[ L(L + 1) +
\al_M \lt(\bt^{(+)*}_M \, {\cal F}^{\rm (G)+}_{L,M} + \bt^{(-)*}_M \, {\cal F}^{\rm (G)-}_{L,M}\rt)\rt] \hb^2 \,
\lt|L, M \rt\rangle
\nn \nn
&  &{} + \bt^{(-)*}_{M+1} \lt[\al_M \, {\cal F}^{\rm (G)+}_{L,M}
+ {1 \over 2} \lt(2 \, M + 1\rt) \, \bt^{(+)}_M\rt] \hb^2 \,
\lt|L, M + 1\rt\rangle
\nn \nn
&  &{} + \bt^{(+)*}_{M-1} \lt[\al_M \, {\cal F}^{\rm (G)-}_{L,M}
- {1 \over 2} \lt(2 \, M - 1\rt) \, \bt^{(-)}_M\rt] \hb^2 \,
\lt|L, M - 1\rt\rangle \, ,
\label{L^2|LM>=}
\ee
which is strictly no longer an eigenvalue equation, though only at second-order in $\bt^{(\pm)}_M$.
By a similar procedure, it can be shown for $\Lvec_z$ that
\be
\Lvec_z \lt|L, M \rt\rangle & = & \lt[ M + {1 \over 2}\lt(\lt|\bt^{(+)}_M\rt|^2 - \lt|\bt^{(-)}_M\rt|^2\rt) \rt.
\nn
&  &{} + \al_M \lt(\bt^{(+)*}_M \, {\cal M}^{\rm (G)+}_{L,M} + \bt^{(-)*}_M \, {\cal M}^{\rm (G)-}_{L,M}\rt)
\nn
&  &{} + \lt. \al^*_M \lt(\bt^{(-)}_M \, {\cal M}^{\rm (G)+}_{L,M - {1 \over 2}}
+ \bt^{(+)}_M \, {\cal M}^{\rm (G)-}_{L,M + {1 \over 2}}\rt)\rt] \hb \,
\lt|L, M \rt\rangle
\nn \nn
&  &{} + \lt[\lt(M + {1 \over 2}\rt) \bt^{(-)*}_{M+1} \, \bt^{(+)}_M +
\al^*_{M+1} \, \bt^{(+)}_M \, {\cal M}^{\rm (G)+}_{L,M + {1 \over 2}} \rt.
\nn
&  &{} + \lt.
\al^*_M \, \bt^{(-)}_{M+1} \, {\cal M}^{\rm (G)+}_{L,M}\rt] \hb \,
\lt|L, M + 1\rt\rangle
\nn \nn
&  & + {} \lt[\lt(M - {1 \over 2}\rt) \bt^{(+)*}_{M-1} \, \bt^{(-)}_M +
\al^*_{M-1} \, \bt^{(-)}_M \, {\cal M}^{\rm (G)-}_{L,M - {1 \over 2}} \rt.
\nn
&  &{} \lt.
+ \al^*_M \, \bt^{(+)}_{M-1} \, {\cal M}^{\rm (G)-}_{L,M}\rt] \hb \,
\lt|L, M - 1\rt\rangle \, ,
\label{Lz|LM>=}
\ee
also no longer an eigenvalue equation at second-order in $\bt^{(\pm)}_M$.

An interesting observation results from considering the diagonal matrix elements for $\Lvec^2$ and $\Lvec_z$.
By supposing that
\be
\lt\langle L, M \rt| \Lvec^2 \lt| L, M \rt\rangle & \approx & L(L + 1) \, \hb^2_{\rm eff.} \, ,
\label{L^2-eigenvalue-effective}
\nl
\lt\langle L, M \rt| \Lvec_z \lt| L, M \rt\rangle & \approx & M \, \hb_{\rm eff.} \, ,
\label{Lz-eigenvalue-effective}
\ee
where $\hb_{\rm eff.}$ becomes a predicted {\em space-time curvature-dependent}
Planck's constant defined with respect to the {\em observed} Planck's constant of
$\hb = 1.054 \, 571 \, 95(07) \times 10^{-34}$~J.s \cite{Robinson} in a flat space-time background,
it is possible to express the gravitational contributions to
(\ref{L^2|LM>=}) and (\ref{Lz|LM>=}) in terms of
\be
\hb_{\rm eff.} & \equiv & \lt(1 + {\Dl \hb_{\rm G}(L,M) \over \hb}\rt) \hb \, ,
\label{Planck-constant-effective=}
\ee
with $\Dl \hb_{\rm G}(L,M)$ a second-order function of $\bt^{(\pm)}_M$.
By relating $\Dl \hb_{\rm G}(L,M)$ to the relative uncertainty in the measurement of Planck's constant,
in the form
\be
{\Dl \hb_{\rm G}(L,M) \over \hb} & < &
\lt. {\Dl \hb \over \hb} \rt|_{\rm expt.} \ \approx \ 6.637 \, 76 \times 10^{-8} \, ,
\label{dh/h=}
\ee
it is possible to establish an upper bound for $\bt^{(\pm)}_M$, such that
$\lt|\bt^{(\pm)}_M\rt|^2 < 10^{-8} \ll 1$.
This observation provides a strong confirmation that, for quantum phenomena
measured in the presence of the Earth's gravitational field, local space-time curvature
has a negligible impact on the observables of orbital angular momentum, as expected,
and provides further justification for adopting the approximations used to
motivate this investigation.

\section{Discussion}
\label{Discussion}

Having established the details of orbital angular momentum for a spinless particle in a locally
curved space-time background, it is useful to examine some of its relevant implications.
An immediate example comes from the recursion relation (\ref{beta-recursion-2b}), showing that
$l = 2$ is the minimum orbital quantum number that accommodates the presence
of local curvature.
A transition to $l = 0$ most likely occurs as a two-step process with the emission
of a single photon at each step, with the understanding that the first transition to
$l = 1$ destroys the boundary conditions necessary to incorporate the
curvature contributions.
As stated earlier, the minimum boundary condition at $l = 2$ coincides with the
spin of a graviton, and suggests that a one-step transition to $l = 0$ is also
theoretically possible, with the simultaneous emission of a single graviton
to conserve angular momentum.
For $l > 2$, it is possible to envision a combination of photon and graviton emissions
that also satisfy known selection rules for each transition.

If graviton emission is reasonable to expect, then conversely such a suggestion also
implies that graviton absorption is possible, in like fashion to photon absorption
under similar conditions.
At the macroscopic level, it is commonly understood that gravitational wave radiation
due to astrophysical sources occurs as freely propagating ripples in space-time,
described quantum mechanically as coherent graviton emissions.
In particular, it is also well-understood that gravitational waves can propagate
to very distant observers without experiencing any noticable dispersion while
passing through matter.
Therefore, at the quantum mechanical level there is an expectation that
coherent emissions must also occur for a many-body quantum system to ultimately
preserve the macroscopic properties of the emitted waveform.
This means that any coherent graviton absorptions for such a system must also translate
into a virtually instantaneous and coherent emission to account for the
anticipated lack of any dispersive effects.

It is no surprise that, given the small local curvature deviation away from an otherwise
flat space-time background, there is no realistic possibility of observing gravitational
contributions to orbital angular momentum under current laboratory conditions.
That is, $\lt|\bt^{(\pm)}_M\rt|^2 \sim 0$ and $\lt|\al_M\rt|^2 \sim 1$ for likely all
relevant experiments performed on Earth.
However, this condition does not necessarily apply when dealing with much stronger
gravitational fields, in which a sufficiently small radius of curvature in the
space-time background exists compared to the particle's orbital radius.
It is conceivable, therefore, that such a situation arises near the event horizon of a
microscopic black hole \cite{Singh-Mobed-Casimir-2}.
In fact, it seems possible that if such a black hole undergoes Hawking radiation,
the backreaction due to time-dependent induced curvature variations
may have a significant impact on the orbital angular momentum of an orbiting
quantum particle nearby, such that corresponding signatures may appear in
photon and graviton emissions corresponding to transitions from
higher to lower levels of $l$.

As a matter of principle, the concept of a curvature-dependent Planck's constant $\hb_{\rm eff.}$
should have an impact on all known physical observables in the quantum domain, such as decay rates
and scattering cross-sections, and any other measurements involving quantized energy states.
In addition, given the interest in trying to observe time-variations in the fine structure constant
over cosmological time scales \cite{Webb}, it may be conceivable to also envision local curvature
variations impacting on spectral line emissions due to matter-energy fluctuations in the early Universe.
Another possibility for identifying the presence of $\hb_{\rm eff.}$ may occur with ultra-high precision
measurements of the gravitational redshift effect, as originally demonstrated by the
Pound-Rebka experiment and its variants \cite{Pound1,Pound2,Vessot}.
In particular, it is conceivable that different theoretical predictions may appear in the
gravitational redshift effect in response to differences in the metric gravitation theories under consideration.
Nevertheless, it remains unrealistic to expect any clearly observable signatures to appear in near-future
measurements of the gravitational redshift effect.


It is important to be reminded that this paper assumes a simply connected topology
for the space-time background.
This necessitates the restriction of an integer-valued spin to describe the
gravitational field \cite{Finkelstein}, if rotated by $2 \pi$ with respect to an
environment described by a larger space-time background in which the field in
question is treated as an isolated system.
The claim by Friedman and Sorkin \cite{Friedman} that it is possible to identify
the existence of a spin-1/2 gravitational background within a nontrivial topological space
is a very interesting observation suggesting that space-time is fundamentally
spinorial in structure.
It is, therefore, worthwhile to consider if a similar analysis of orbital
angular momentum within a nontrivial topological setting leads to the same
conclusions as first drawn by Friedman and Sorkin, but based upon a
completely different set of motivations.
Given this paper's suggestion that the gravitational part of
the orbital angular momentum operator is sensitive to half-integer spacings
along the axis of quantization, it seems reasonable to suggest that a
pre-existent spin-1/2 structure is somehow embedded within the restrictions
imposed by a simply connected topology, due to the boundary conditions
that define $\al_M$ and $\bt^{(\pm)}_M$.
Within the context of quantum gravity, this is an important issue to consider
if Wheeler's ``space-time foam'' concept \cite{Wheeler} is well-posed,
such that topologically dynamical structures spontaneously emerge at a
sufficiently small length scale.
In principle, such effects may be reflected within $\al_M$ and $\bt^{(\pm)}_M$
as time-dependent signatures with potentially large variations in their
relative magnitudes, though the likelihood of envisioning an experiment
to test this hypothesis seems remote.

Finally, a different type of consideration to follow from this paper involves the
coupling of rotation $\Omvec$ to the orbital and spin angular momentum of a quantum particle.
Recently, it is shown by Shen \cite{Shen} that Mashhoon's proposed
spin-rotation coupling interaction $\Omvec \cdot \Svec$ \cite{Mashhoon-SR},
introduced as an extension of the hypothesis of locality for a quantum system,
can be generalized to describe the coupling of graviton spin $\Svec^{\rm (G)}$
to a gravitomagnetic field identified with $\Omvec$.
This results in a gravitational self-interaction contained within the
space-time background.
A subsequent computation by Ramos and Mashhoon \cite{Ramos} shows how this
generalization is applicable for the propagation of gravitational waves,
while Papini \cite{Papini} develops the approach further to accommodate massive
spin-2 particles and explore its application to wave optics phenomena, such
as gravitational lenses.
In light of Shen's generalization of the Mashhoon effect,
it may be possible to interpret the gravitational orbital angular momentum operator
$\Lvec^{\rm (G)}$ as an analogous modification of the orbital-rotation
coupling term to accompany the spin-rotation coupling generalization.
In like fashion, this interaction results in a generalization
of the Sagnac effect $\Omvec \cdot \Lvec$, such that for a spinless particle,
\be
\Omvec \cdot \Jvec & = & \Omvec \cdot \lt(\Jvec^{(0)} + \Jvec^{\rm (G)}\rt)
\label{omega.J=}
\ee
gives rise to an overall phase shift in the wavefunction generated by
the Sagnac and Mashhoon effects \cite{Mashhoon-SR1},
where $\Jvec^{(0)} = \Lvec^{(0)}$, and
$\Jvec^{\rm (G)} = \Lvec^{\rm (G)} + \Svec^{\rm (G)}$.
It remains to be seen whether these type of quantum mechanical predictions
involving classical gravitation are observable with current or near-future
experimental means available.

\section{Conclusion}
\label{Conclusion}

This paper demonstrates the impact of local space-time curvature on the orbital angular momentum of a spinless particle
in quantum mechanics.
It suggests the existence of half-integer spacings along the axis of quantization that are otherwise not accessed
in the absence of gravitational interactions with quantum matter.
In addition, the constraints required to preserve consistency of the formalism demonstrates
that $l = 2$ is the minimum allowable value for the orbital quantum number to incorporate curvature
contributions, which precisely coincides with the spin of a graviton in linearized general relativity.
The consequences of these details are compared with previous research that suggests the existence
of a spin-1/2 internal structure embedded in space-time, with interesting consequences for
potentially advancing quantum gravity research into the future.

Possible future research may involve studying the addition of both orbital and spin angular momentum in
the presence of a local gravitational background, to determine if curvature-dependent quantities make
a contribution to the Clebsch-Gordan coefficients.
Other possibilities may include the incorporation of non-trivial topological backgrounds to see
if a spin-1/2 gravitational background alters the constraint equations required to specify the
gravitational ladder coefficients.
It is also worthwhile to determine if these observations of orbital angular momentum get adequately reflected
within a quantum field theory framework.
These and other possibilities may be investigated in due course.

\section{Acknowledgements}

The authors wish to thank Bahram Mashhoon and Hubert de Guise for helpful comments and suggestions concerning this paper.
They also acknowledge financial support from the University of Regina, Faculty of Science.

\end{document}